\documentclass[prx,aps,twocolumn,superscriptaddress]{revtex4-1}

\usepackage[pdftex]{graphicx}
\usepackage{amsbsy,amssymb,amsmath,bm,mathtools}

\usepackage{xspace}
\usepackage{bm}
\usepackage{color}
 
\usepackage{url}
\usepackage[section]{placeins}
\usepackage[colorlinks=true,linkcolor=blue,citecolor=blue]{hyperref}

\begin{document}


\title{Nonequilibrium generation of charge defects in kagome spin ice under slow cooling}

\author{Zhijie Fan}
\affiliation{Department of Physics, University of Virginia, Charlottesville, VA 22904, USA}
\affiliation{University of Science and Technology of China}
\affiliation{Hefei National Laboratory for Physical Sciences at the Microscale}

\author{Gia-Wei Chern}
\affiliation{Department of Physics, University of Virginia, Charlottesville, VA 22904, USA}

\date{\today}

\begin{abstract}
Kagome spin ice is one of the canonical examples of highly frustrated magnets. The effective magnetic degrees of freedom in kagome spin ice are Ising spins residing on a two-dimensional network of corner-sharing triangles. Due to strong geometrical frustration, nearest-neighbor antiferromagnetic interactions on the kagome lattice give rise to a macroscopic number of degenerate classical ground states characterized by ice rules. Elementary excitations at low temperatures are defect-triangles that violate the ice rules and carry an additional net magnetic charge relative to the background. We perform large-scale Glauber dynamics simulations to study the nonequilibrium dynamics of kagome ice under slow cooling. We show that the density of residual charge defects exhibits a power law dependence on the quench rate for the class of algebraic cooling protocols. The numerical results are well captured by the rate equation for the charge defects based on the reaction kinetics theory. As the relaxation time of the kagome ice phase remains finite, there is no dynamical freezing as in the Kibble-Zurek scenario. Instead, we show that the power-law behavior originates from a thermal excitation that decay algebraically with time at the late stage of the cooling schedule.  Similarities and differences in quench dynamics of other spin ice systems are also discussed.  
\end{abstract}

\maketitle

\section{Introduction}
\label{sec:intro}

The nonequilibrium dynamics of many-body systems following a quench have been extensively studied over the years. Several universal behaviors after a fast quench have been established, which depend on the symmetry and conservation law of the order parameter field~\cite{bray94,puri09,onuki02}. The kinetics of phase ordering following a fast quench is often governed by annihilation dynamics of  topological defects in symmetry-breaking phases. On the other hand, for systems which are slowly quenched across a critical point, the Kibble-Zurek (KZ) mechanism  offers a general framework for nonequilibrium dynamics and the formation of topological defects~\cite{kibble76,kibble80,zurek85,zurek93}. In particular, it shows that the excess defects left at the end of the cooling falls off with the annealing rate according to a power law whose exponent is determined by the equilibrium critical properties~\cite{decampo14}. Originally developed to understand the density of relic topological defects in early universe, the KZM has since been confirmed in phase transition dynamics of various condensed matter systems~\cite{zurek93,decampo14}.  

The central idea of KZ mechanism is the breaking of adiabaticity due to critical slowing down when approaching a phase transition point. Specifically, consider a general cooling schedule that starts at $t=0$ and reaches the critical point at $t = \tau_Q$. For thermal quenches across a critical point at $T_c$, the control parameter is given by $\epsilon(t) = (T_c - T(t)) / T_c$. The relaxation time $\tau$ of the system is now time-dependent through its dependence on $\epsilon(t)$.  When the relaxation time $\tau(t)$ is shorter than the time-scale $\epsilon/\dot{\epsilon}$ that characterizes the change of the control parameter, quasi-equilibrium can be quickly established and adiabaticity is maintained.  The condition $\tau(t^*) = \epsilon/\dot{\epsilon}$ thus determines a freeze-out time~$t^*$ which signals the onset of non-adiabaticity. In the limit of slow cooling, a universal scaling relation for the freeze-out time $\hat{t} = \tau_Q - t^*$ (measured from the critical point) is obtained $\hat{t} \sim \tau_Q^{\nu z/(1 + \nu z)}$, where $\nu$ is the exponent characterizing the power-law divergence of equilibrium correlation length $\xi \sim |T-T_c|^{-\nu}$, and $z$ is the dynamical exponent $\tau \sim \xi^z$. Importantly, the residual density of topological defects can be estimated by the correlation length at the freeze-out time:
\begin{eqnarray}
	n_d \sim \hat{\xi}^{-D} \sim \tau_Q^{-D \nu/(1+ \nu z)},
\end{eqnarray}
where $D$ is the spatial dimension of the system. 

The KZ mechanism can also be generalized to describe nonequilibrium dynamics of the paradigmatic 1D Ising ferromagnet which exhibits an unconventional critical point at $T_c = 0$~\cite{suzuki09,krapivsky10,jeong20,priyanka21,mayo21}. Topological defects are kinks and anti-kinks that connect the doubly degenerate fully polarized ground states. Contrary to standard continuous phase transitions, both correlation length and relaxation time of the Ising chain diverge exponentially as $T \to 0$. Yet, a dynamical exponent can still be defined, e.g. $z = 2$ for Glauber dynamics, and a freeze-out time $\hat{t}$ can be determined from the KZ condition $\tau(\tau_Q - \hat{t}) = \hat{t}$. However, while scaling relation is still obtained for the residual defect density, the exponent of the resultant power law depends on the cooling schedule. 

\begin{figure}[t]
\centering
\includegraphics[width=0.99\columnwidth]{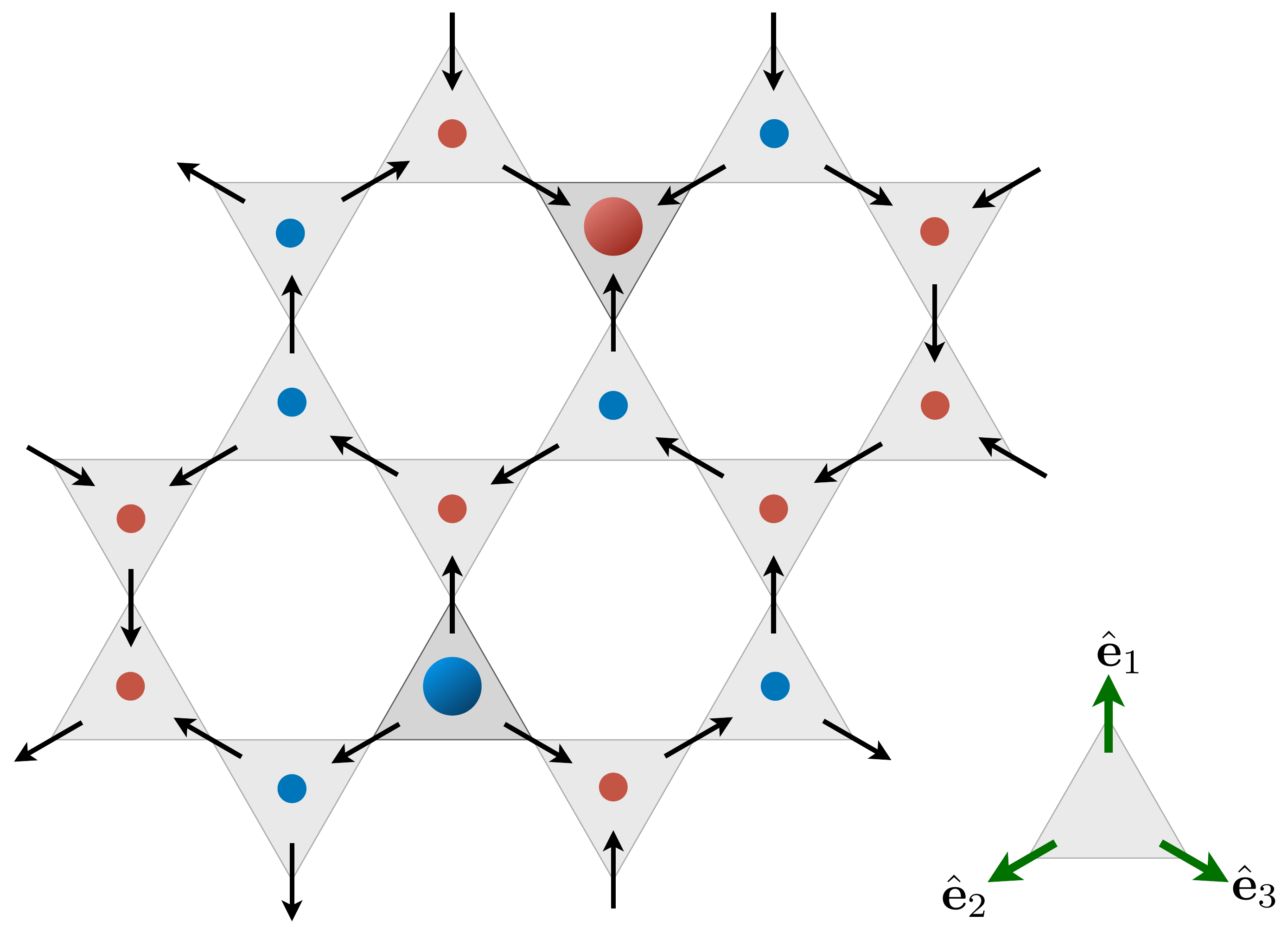}
\caption{Schematic of the kagome spin ice. The unit vectors $\hat{\mathbf e}_{1,2,3}$ denote the easy-axis directions of the three sublattices. The small red and blue circles, indicating a magnetic charge $Q = +1$ and $-1$, respectively, are triangles that satisfy ice rule constraints. The bigger red and blue circles represent defect triangles with a net magnetic charge of $+3$ and $-3$, respectively. }
    \label{fig:kagome-ice}
\end{figure}

It is worth noting that the defect formation in the KZ scenario is intimately related to broken symmetries. As the order parameter in the ordered phase can take on multiple values due to global symmetry of the system, the order parameter in general cannot be the same in regions which are beyond the equilibrium correlation length. The incompatibility of different ordered domains gives rise to topological defects, such as kinks or vortices, which are localized regions that connect two or more adjoining domains of different order parameters.  A different kind of localized defects occur in spin ices~\cite{harris97,ramirez99,bramwell01,spin-ice-book} and similar highly constrained systems~\cite{hfm-book,baxter08}. Due to strong geometrical frustration, spin ice remains in a disordered state at temperatures well below the dominant exchange energy scale. Yet, contrary to the uncorrelated paramagnets at high temperatures, spins in the low-$T$ ice phase are strongly correlated. The short-range correlation is dictated by local constraints, known as the ice rules, defined on local simplex such as triangle in kagome ice; see FIG.~\ref{fig:kagome-ice}. Importantly, violation of the ice rules gives rise to defect simplexes, which are particle-like elementary excitations of the ice phase.

In this paper, we study the nonequilibrium dynamics of excess charge defects in short-range kagome spin ice under slow quenches. Kagome spin ice is effectively an antiferromagnetic Ising model defined on the kagome lattice, a two-dimensional network of corner-sharing triangles as shown in FIG.~\ref{fig:kagome-ice}. We perform extensive Glauber-dynamics Monte Carlo simulations of the annealing process of kagome ice. In particular, for a class of algebraic cooling schedules, the residual defects are shown to exhibit a power-law scaling with respect to the cooling rate, with an exponent dependent on the exponent of the algebraic cooling. We further adopt the reaction kinetics theory to describe the transition kinetics of triangle-simplexes in kagome ice. The results from the integration of the rate equations agree excellently with the Glauber dynamics simulations.

Contrary to standard KZ scenarios, the kagome spin ice does not exhibit a critical state at any finite or zero temperature. In fact, the relaxation time is found to be nearly temperature independent in the ice phase. Excess defects originate from the time-varying thermal generation against a constant decay. As a result, there is no significant freezing in the annealing process. Instead, the defect density at the end of cooling results from the accumulation of excess defects within the relaxation time. This process is well captured by an analytical formula obtained from the rate equation, which gives rise to power-law scaling of residual defect density in the case of algebraic cooling protocol.

The rest of the paper is organized as follows. In Sec.~\ref{sec:glauber}, we present details of the Glauber dynamics Monte Carlo (MC) simulations for kagome spin ice and introduce parameters for algebraic cooling schedules. In Sec.~\ref{sec:kinetics}, we adapt the chemical reaction theory to describe the transition kinetics of triangle-simplexes in kagome ice and derive a rate equation for the dynamical evolution of the defect triangles. An asymptotic analytical solution for the scaling behavior in the slow cooling limit is presented in Sec.~\ref{sec:sol}. Finally, a summary and outlook is presented in Sec.~\ref{sec:summary}.

\section{Glauber dynamics simulations of kagome spin ice}
\label{sec:glauber}

Spin ices are an unusual class of frustrated ferromagnets where strong easy-axis anisotropy results in an effective antiferromagnetic Ising model defined on a lattice with corner-sharing simplexes. For kagome spin ice, the building block simplex is triangle, and the local easy-axis is along the line that connects two adjacent triangles. We introduce unit vectors $\hat{\mathbf e}_i$ to specify the easy-axis direction, which points from the center of the up-triangle to the corners. The magnetic moments are then expressed as $\mathbf S_i = \sigma_i \mu_0 \hat{\mathbf e}_i$, where $\mu_0$ is the magnitude of the magnetic moment, $\hat{\mathbf e}_i$ is the local crystal-field axis, and the Ising variable $\sigma_i = \pm 1$ indicates the direction of the magnetic moment. 

In this work, we consider a kagome spin ice model with interactions restricted to the nearest neighbors, originally studied by Wills {\em et al.}~\cite{wills02}. Effects due to long-range dipolar interactions will be discussed in Sec.~\ref{sec:summary}. The geometry of the lattice is such that $\hat{\mathbf e}_i \cdot \hat{\mathbf e}_j = -1/2$ for any nearest-neighbor pair $\langle ij \rangle$. As a result, a ferromagnetic exchange interaction between nearest-neighbor pairs $J_F \, \mathbf S_i \cdot \mathbf S_j$, with $J_F < 0$, gives rise to an antiferromagnetic Ising model on the kagome lattice
\begin{eqnarray}
	\label{eq:H-spinice}
	\mathcal{H} = J \sum_{\langle ij \rangle} \sigma_i \sigma_j = \frac{v}{2}\sum_{\alpha} Q_\alpha^2 + E_0,
\end{eqnarray}
where $J = |J_F|/2$ is the effective interaction strength. In the second equality above, the Hamiltonian is expressed in terms of magnetic charges $Q_\alpha$ associated with triangles. It is convenient to introduce the dumbbell representation of spin ice~\cite{castelnovo08}, in which magnetic dipoles are stretched into bar magnets of length $\ell$ such that their poles meet at the centers of triangles. The magnetic charge associated with each pole is then $q_m = \mu_0/\ell$. One can then define a total magnetic charge for each triangle:
\begin{eqnarray}
	Q_\alpha = \pm q_m \sum_{i \in \alpha} \sigma_i, 
\end{eqnarray}
where $+$ and $-$ signs are used for up- and down-triangles, respectively. The self-energy coefficient in Eq.~(\ref{eq:H-spinice}) is given by $v = J \ell^2/\mu_0^2$ . In the following, the magnetic charges of triangles are expressed using $q_m$ as the unit. 

The ground states of the spin ice Hamiltonian Eq.~(\ref{eq:H-spinice}) are then given by states with minimum magnetic charges on every triangle. As there are three Ising spins in a triangle, the minimum charge condition is thus $Q_\alpha = +1$ or $-1$, which correspond to 2-in-1-out or 1-in-2-out, respectively, spin configurations. These local constraints for the ground states are known as ice rules based on the analogy with the Bernal-Fowler rules for water ice~\cite{bernal33}. The number of ground states satisfying these local constraints increases exponentially with the number of spins, giving rise to a finite residual entropy even at $T = 0$. The residual entropy  $S_0/N = \frac{1}{3} \ln\frac{9}{2} = 0.501$ estimated by Pauling's method agrees very well with Monte Carlo simulations~\cite{wills02}.
Importantly, the magnetic charge expression for the Hamiltonian in Eq.~(\ref{eq:H-spinice}) also shows that elementary excitations in the degenerate ground-state manifold are defect triangles with a net charge $Q = +3$ or $-3$, corresponding to 3-in or 3-out spin configurations, respectively. 

It is worth contrasting the degenerate ice manifold of kagome with that of the pyrochlore (or its 2D checkerboard analog) spin ice. The basic simplex unit in pyrochlore is a tetrahedron with four spins. As a result, a similar minimum charge $Q = 0$ leads to the well-known 2-in-2-out ice rules. Elementary excitations of the low temperature ice phase are tetrahedra with $Q = \pm 2$, which behaves as emergent magnetic monopoles in a charge-free vacuum~\cite{jaubert09}. This difference in ice rules and elementary excitations also lead to fundamentally different quenched-induced nonequilibrium dynamics, which will be briefly discussed in Sec.~\ref{sec:summary}.

Monte Carlo method with Glauber dynamics for Ising spins~\cite{glauber63,krapivsky-book} is employed to study the quench dynamics of kagome spin ice. At every time step, one random spin, say at site-$i$, is updated according to the transition probability 
\begin{eqnarray}
	\label{eq:glauber}
	w(\sigma_i \to -\sigma_i; t) = \frac{1}{2} \left(1 - \tanh\frac{\beta(t) \Delta E_i}{2} \right)
\end{eqnarray}
where $\beta(t) = 1/T(t)$ is the time-dependent inverse temperature, and $\Delta E_i$ is the energy change due to a flipped~$\sigma_i$. For a cooling schedule from $t = 0$ to $t = \tau_Q$, the simulation time is increased by $\delta t = 1/N_s$ after each attempt of single-spin update; here $N_s = 3 L^2$ is the total number of spins. A time-step $\Delta t = 1$ thus corresponds to $N_s$ single-spin updates. All MC simulations presented in the following were performed with a linear size $L = 100$, corresponding to $N_s = 30,000$ spins. 

There are three types of single-spin update in kagome ice: (i) $\Delta E_i = 0$ corresponding to an exchange of magnetic changes on the two triangles sharing~$\sigma_i$. (ii) $\Delta E_i = \pm 4J$ due to the creation/annihilation of a single $Q = \pm 3$ defect, and (iii) $\Delta E_i = \pm 8J$ caused by creation/annihilation of a pair of adjacent defect-triangles of opposite charges. The reaction dynamics corresponding to these spin updates are discussed in Sec.~\ref{sec:kinetics} below. At low temperatures, transitions that cost energy are dominated by the creation of single defect-triangle with $\Delta E = 4J$. For Glauber dynamics Eq.~(\ref{eq:glauber}), this suggests a dimensionless parameter 
\begin{eqnarray}
	\label{eq:gm-factor}
	\gamma(t) = \tanh[2 \beta(t) J], 
\end{eqnarray}
which is similar to the one introduced for quench dynamics of 1D Ising chain~\cite{krapivsky10}. The resultant transition rate for such energy-costly process is $w(t) = \frac{1}{2}[1-\gamma(t)]$.

\begin{figure}[t]
\includegraphics[width=0.85\linewidth]{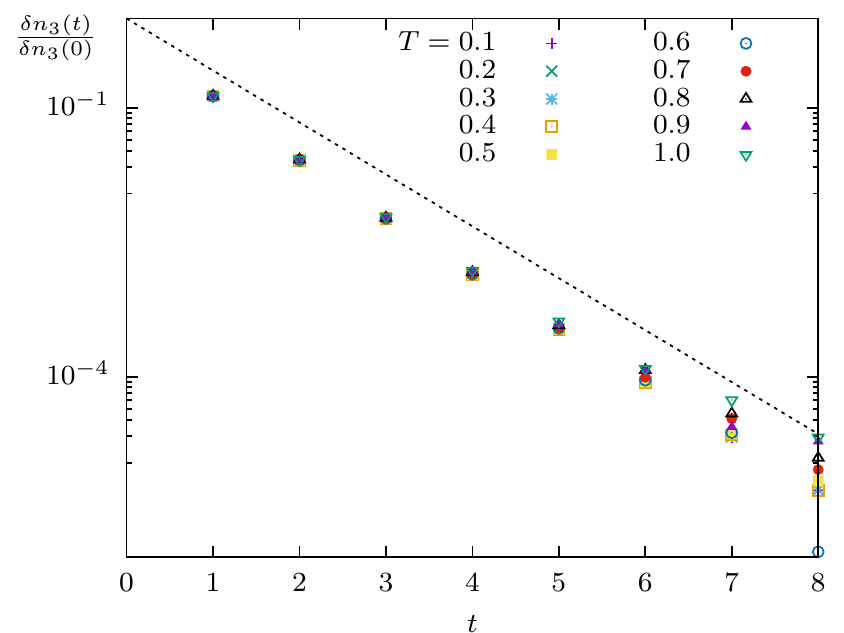}
\caption{Semi-log plot of the normalized density of excess charge defect $\delta n_3$ versus time for kagome ice under instant quench to various temperatures. The density is normalized by the initial value of random spin configuration where the fraction of 3-in/3-out triangles is $n_3(0) = 1/4$. The dashed line shows an exponential decay $e^{-t/\tau}$ with $\tau \approx 0.75$.}
    \label{fig:relaxation}
\end{figure}

The Glauber dynamics MC method is employed to investigate the temperature dependence of relaxation time $\tau(T)$ for kagome spin ice. To this end, we perform thermal quench simulations where the system is suddenly quenched from a random state to a constant low temperature at $t = 0$. We use $n_Q$ to denote the density of triangles with charge $Q$. The defect density is given by $n_3 = (n_{+3} + n_{-3})$. The charge neutrality $n_{+3} = n_{-3}$ is preserved during the relaxation process. FIG.~\ref{fig:relaxation} shows the decay of the density of excess defects, $\delta n_3(t) = n_3(t) - n_3^{\rm eq}(T)$, after the quench. The resultant curves of different temperatures, after normalization by the initial value, fall nearly on the same straight line in the semi-log plot, indicating an exponential decay
\begin{eqnarray}
	\label{eq:tau}
	\delta n_3(t) = \delta n_3(0)\,e^{-t/\tau(T)}.
\end{eqnarray}
The estimated relaxation time is $\tau \approx 0.75$, which represents an estimate of the decay constant averaged over all temperatures. As shown by the spreading of data points at large time (e.g. $t = 8$), the relaxation time is weakly temperature dependent, with a smaller value at higher temperatures. As will be discussed in Sec.~\ref{sec:kinetics}, this weak temperature dependence can be attributed to the interplay between the two reaction pathways involving defect triangles: the spontaneous decay and the pair annihilation.

\begin{figure}[t]
\includegraphics[width=0.85\linewidth]{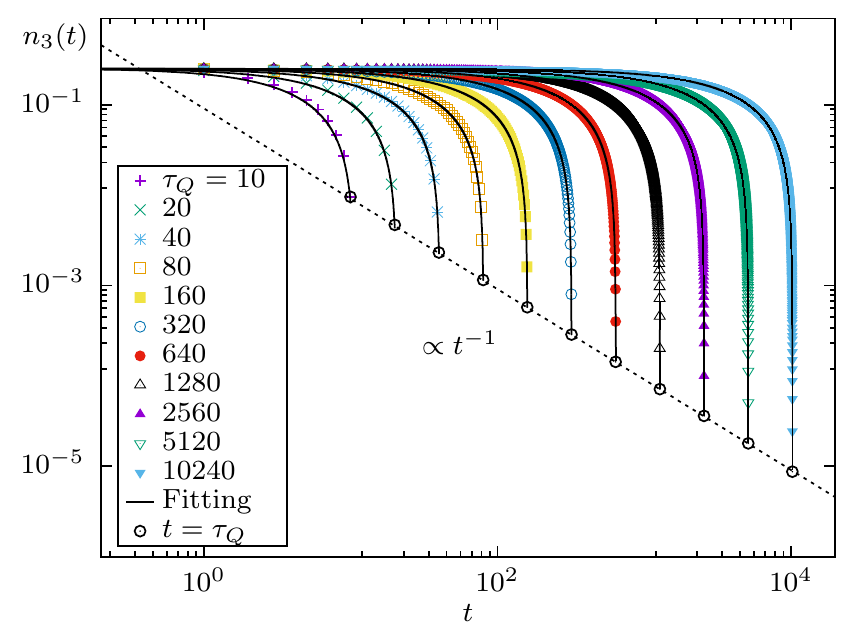}
\includegraphics[width=0.85\linewidth]{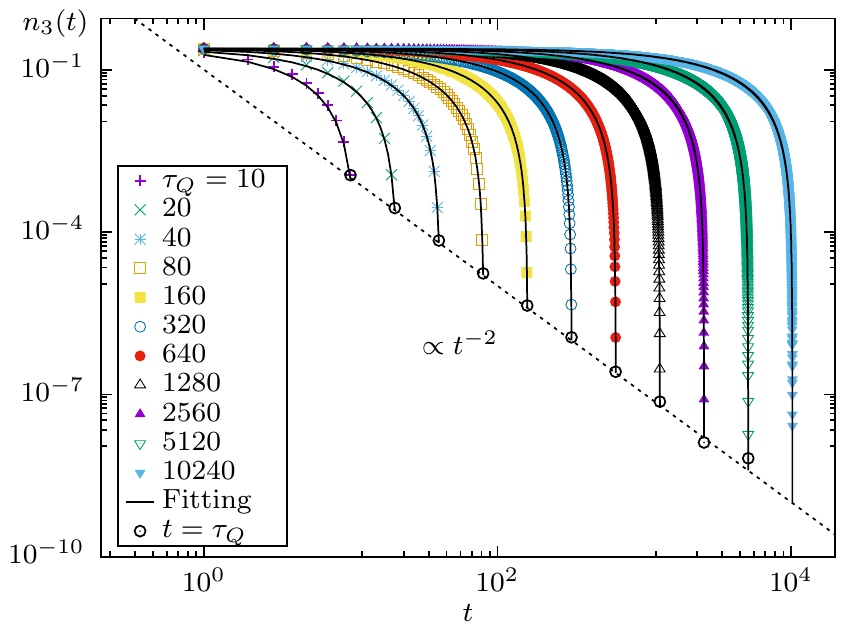}
\includegraphics[width=0.85\linewidth]{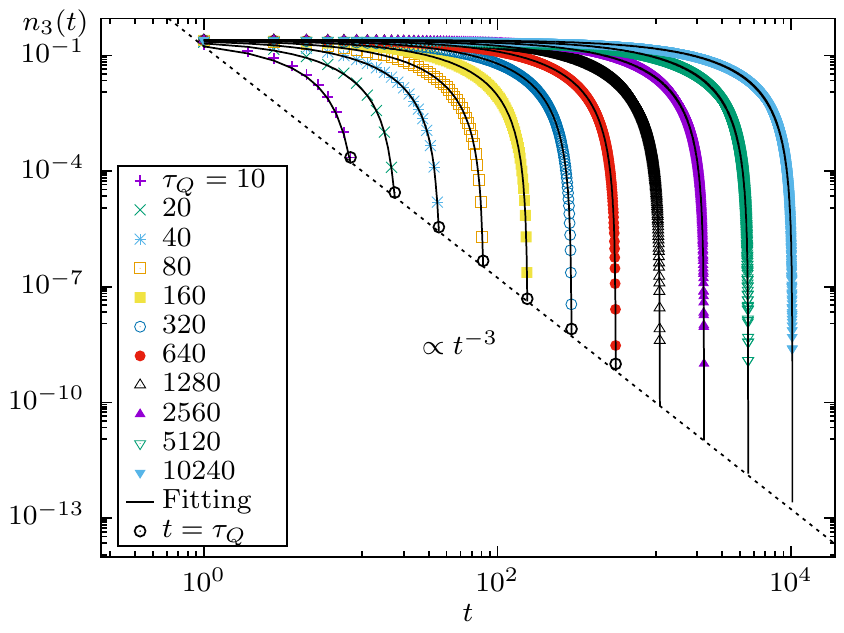}
\caption{Relaxation of charge defect density $n_3(t)$ as function of time $t$ for the kagome spin ice undergoing algebraic cooling schedules with (a) $\alpha = 1$ (``linear" cooling), (b) $\alpha = 2$, and (c)~$\alpha = 3$ with different values of the quench time $\tau_Q$. The solid black curves are fitted with mean-field results with parameters $A_1=1.566$ and $A_2=0.600$. The dashed lines correspond to the analytical formula Eq.~(\ref{eq:n3-analytical}) obtained from the asymptotic solution of the rate equation. The final defect density $n_3(t=\tau_Q)$ (empty black circles) exhibits power-law relaxation with cooling time $\tau_Q$. }
\label{fig:nd-cooling}
\end{figure}

The fact that the relaxation of charge defects is well described by an exponential decay also indicates their non-topological nature. The dominant process of the relaxation is through the decay of, e.g. a $Q = +3$ defect into a $+1$ triangle, while converting a neighboring $Q = -1$ triangle to $+1$. Pair annihilation of $\pm 3$ defects are negligible in the equilibration. This is in stark contrast to the case of pyrochlore spin ice where pair annihilation of $Q = \pm 2$ magnetic monopoles are the driving process of relaxation after quench, giving rise to a $1/t$ decay of excess monopole density when quenched to zero temperature~\cite{castelnovo10}.

Next we apply the Glauber dynamics simulations to study relaxation of charge defects under slow cooling. To this end, we consider a class of algebraic cooling schedules, similar to the ones introduced in Ref.~\cite{krapivsky10} for the quench dynamics of Ising chain. In terms of the dimensionless parameter $\gamma(t)$, the thermal-bath temperature varies with time according to 
\begin{eqnarray}
	\label{eq:gamma}
	1-\gamma(t) = \left(1 - \frac{t}{\tau_Q} \right)^\alpha.
\end{eqnarray}
In particular, the case of $\alpha = 1$ corresponds to linear increase of $\gamma$ from zero at $t = 0$ to $\gamma = 1$ at the end of cooling. Regardless of the exponent $\alpha$, the physical temperature goes from $T(t=0) = \infty$ to zero at $t = \tau_Q$.

For a given exponent $\alpha$ of the algebraic cooling schedule, Glauber dynamics simulations were performed with the cooling time $\tau_Q = 10\times 2^m$, where $m=0, 1, 2, \dots, 10$. FIG.~\ref{fig:nd-cooling} shows the defect density $n_3(t)$ versus time for three different exponents $\alpha = 1, 2,$ and 3 and varying cooling time $\tau_Q$. For $m \le 7$, each data point was obtained by averaging results from more than $10,000$ randomly generated initial states, while for $m > 7$, more than $1500$ randomly generated initial states are averaged. The overall behavior is that of a slow decrease of charge defects for most of relaxation process, followed by a rather rapid drop at the end of cooling. Importantly, the residual defect density left at $t = \tau_Q$ exhibits a power-law dependence on the cooling time
\begin{eqnarray}
	n_3(\tau_Q) \sim \tau_Q^{-\mu},
\end{eqnarray}
where the exponent $\mu$ seems to be given by the exponent $\alpha$ of the algebraic cooling. These results are rather intriguing since, as demonstrated above, the relaxation time $\tau(T)$ remains rather short. Unlike 1D Ising chain or pyrochlore spin ice, the nearest-neighbor kagome spin ice does not exhibit a critical point at $T  = 0$. As a result, the system does not suffer from KZ-type freezing due to critical slowing down and finite cooling rate.

\section{Reaction kinetics of charge defects}
\label{sec:kinetics}

\begin{figure}[t]
\centering
\includegraphics[width=0.99\columnwidth]{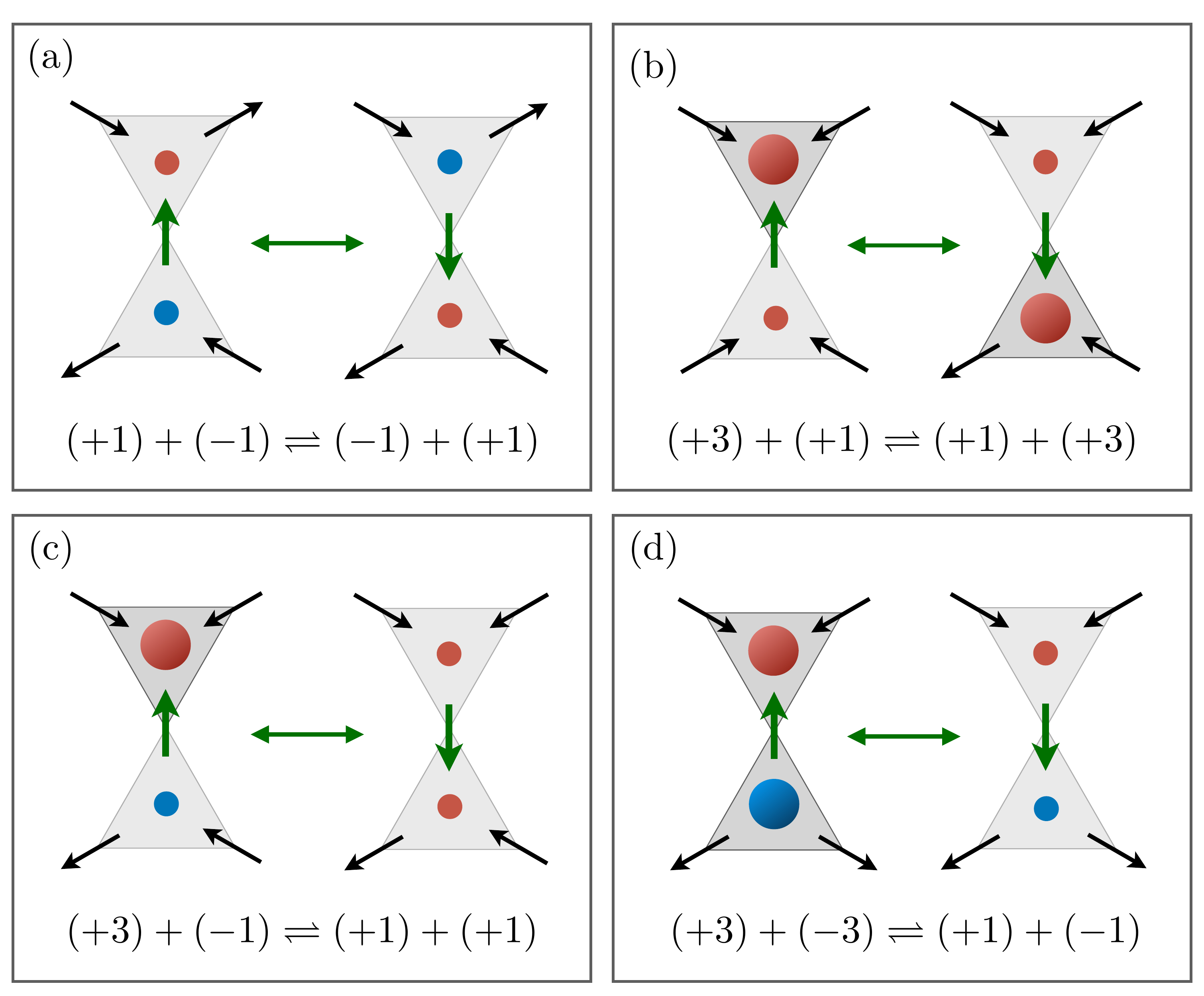}
\caption{Summary of simplex reactions due to a single spin-flip in kagome spin ice. (a) exchange of $+1$ and $-1$ charges. (b) exchange of $+3$ and $+1$ charges (and similar one for $-3$ and $-1$). (c) reaction of $+3$ and $-1$ into two $+1$ triangles (and the time-reversal counterpart). (d) pair-annihilation of $Q=\pm3$ charges. }
\label{fig:reactions}
\end{figure}

To understand the scaling behaviors of the charge defects in kagome spin ice, we adopt the reaction kinetics theory to describe the dynamical evolution of the defect triangles. The basic idea is to describe the time evolution in terms of number densities of triangle-simplexes of different charges in a mean-field sense. For convenience, we also borrow terms from chemical reaction theory and use species to refer to triangles of different charges. Assuming that the magnet remains spatially homogeneous during relaxation, rate equations are employed to describe the ``chemical reactions" of different triangle species. In kagome spin ice, there are 4 different spices, corresponding to triangles with magnetic charge $Q = \pm 1$, and $\pm 3$. In terms of magnetic charges, FIG.~\ref{fig:reactions} summarizes the four distinct types of processes due to a single spin-flip. The first one in panel~(a) shows the exchange of $+1$ and $-1$ charges between two adjacent triangles. The process shown in FIG.~\ref{fig:reactions}(b) describes the exchange of $+3$ and $+1$ changes between nearest-neighbor triangles, which can be thought of as the diffusion of the $+3$ defect. In both cases, the energy is conserved, $\Delta E = 0$, and there is no net change in simplex spices. As a result, these two processes are similar to ``physical process", as opposed to the chemical processes to be discussed below.

The process depicted in FIG.~\ref{fig:reactions}(c) corresponds to the decay of a $+3$ defect into a $+1$ triangle, while shedding the extra charge to convert a neighboring $-1$ triangle into $+1$. The resultant ``chemical" reaction and its reverse can be summarized as 
\begin{eqnarray}
	\label{eq:3-decay}
	(+3) + (-1) \rightleftharpoons (+1) + (+1).
\end{eqnarray} 
And finally, FIG.~\ref{fig:reactions}(d) describes the pair-annihilation of $\pm 3$ charge defects into $\pm1$ ice-rules obeying triangles, with the chemical reactions: 
\begin{eqnarray}
	\label{eq:pair-annihilation}
	(+3) +(-3) \rightleftharpoons (+1)+(-1).
\end{eqnarray} 
It is worth noting that since these reactions are induced by the flipping of a magnetic dipole, the total magnetic charge is always conserved. 

Next we use chemical reaction theory to derive constraints on the reaction rates, which are required for obtaining the rate equations. To this end, we consider a general chemical reaction
\begin{eqnarray}
	\label{eq:Q-reaction}
	Q_A + Q_B \rightleftharpoons Q_C + Q_D,
\end{eqnarray}
where $Q_A, ~Q_B$ are the initial reactants, and $Q_C, ~Q_D$ are the final products. Charge conservation requires $Q_A + Q_B = Q_C + Q_D$. The two-way harpoon indicates that the reaction can occur in both forward and reversed directions.
It is convenient to choose the forward direction as the one that lowers the total energy, i.e., $\Delta E < 0$. These are consistent with both reactions in Eqs.~(\ref{eq:3-decay}) and~(\ref{eq:pair-annihilation}). In other words, the forward reaction is the decay or the annihilation of magnetic charges, while the reversed reaction is the excitation of magnetic charges. 


For a given reaction, its rate is proportional to densities of the reactants. For example, the transition rate of forward reaction for Eq.~(\ref{eq:Q-reaction}) is $v_+ \propto n_{Q_A} n_{Q_B}$. The net rate of reaction in the forward direction is then
\begin{align}
    v = v_{+} - v_{-} = k_+ n_{Q_A} n_{Q_B} - k_{-1}n_{Q_C} n_{Q_D},
\end{align}
where $n_Q$ is the density of triangles with charge $Q$, and $k_{\pm}$ denote the reaction coefficients of forward/reversed reactions, respectively. 
These reaction coefficients, however, are not independent. When the system reaches equilibrium, the net change is zero $v = 0$, which in turn means $k_+/k_- = n^{\rm eq}_{Q_C} n^{\rm eq}_{Q_D} / n^{\rm eq}_{Q_A} n^{\rm eq}_{Q_B}$.
The equilibrium densities of the various species are given by the Boltzmann distribution, $n^{\rm eq}_Q = g_Q \,e^{-\beta E_Q} / Z$, where $Z$ is the partition function, $E_Q$ is the energy of charge species $Q$, and $g_Q$ is its degeneracy. We thus have
\begin{eqnarray}
	\label{eq:ratio_k}
	\frac{k_+}{k_-} =  \frac{g_{Q_C} g_{Q_D} }{g_{Q_A} g_{Q_B}}e^{-\beta \Delta E},
\end{eqnarray}
where $\Delta E$ is the energy difference between products and reactants. In general, the reaction coefficients $k_\pm$ can be expressed as
\begin{eqnarray}
	\label{eq:e_av}
	k_{\pm} = A_{\pm} e^{-\beta \varepsilon_{\pm}},
\end{eqnarray}
where $\varepsilon_{\pm}$ are the activation energies for the forward/backward reactions, respectively. In chemical reactions which often involve an intermediate state, these energy barriers are the energy differences between the intermediate state and the initial/final state, respectively. The coefficients $A_{\pm}$ are now nearly temperature independent. Let $E^*$ be the energy of the intermediate state, we have $\varepsilon_+ = E^* - (E_{Q_A} + E_{Q_B})$ and $\varepsilon_- = E^* - (E_{Q_C} + E_{Q_D})$. Substitute Eq.~(\ref{eq:e_av}) into the ration in Eq.~(\ref{eq:ratio_k}), and using the fact that $\varepsilon_+ - \varepsilon_- = \Delta E$, we obtain the ratio between the two pre-factors
\begin{eqnarray}
	\frac{A_+}{A_-} =  \frac{g_{Q_C} g_{Q_D} }{g_{Q_A} g_{Q_B}}.
\end{eqnarray} 
The overall reaction rate, and in particular its temperature dependence, naturally also depends on the energy level $E^*$ of the intermediate state. However, for Ising spins with Glauber dynamics, the transition rate Eq.~(\ref{eq:glauber}) only depends on the energy difference $\Delta E$, which does not involve any intermediate state. Or equivalently, the initial state with a higher energy serves as such intermediate, hence $\varepsilon_+=0$ and $\varepsilon_- = |\Delta E|$.

With these simplifications, there is only one independent parameter, e.g. $A_-$, for the determination of the net reaction rate
\begin{align}
    v = A_{-} \left[ \frac{g_{Q_C} g_{Q_D}}{g_{Q_A} g_{Q_B}} n_{Q_A} n_{Q_B} - e^{-\beta \left|\Delta E\right|} n_{Q_C} n_{Q_D} \right].
    \label{eq::spin_ice_reaction_rate}
\end{align}
When a charge species is involved in multiple reactions, the rate equation of its density should include contributions from all possible reactions. For the case of defect triangles with $Q = \pm 3$, one needs to consider both the decay and pair-annihilation. As a result, we have $dn_{+3}/dt = -v_{\rm decay} - v_{\rm pair}$, where $v_{\rm decay}$ and $v_{\rm pair}$ are the reaction rates of Eqs.~(\ref{eq:3-decay}) and~(\ref{eq:pair-annihilation}), respectively. The rate equations can be further simplified by taking into account the charge neutrality conditions, $n_{+3} = n_{-3}$ and $n_{+1} = n_{-1}$, which are verified to high precision in our Glauber dynamics. We define the total density of charge defects as $n_3 = (n_{+3} + n_{-3})$, and the density of the ground-state triangles as $n_1 = (n_{+1} + n_{-1})$. As the total number of triangles in a kagome lattice is fixed, we have $n_1 = 1 - n_3$ and only need to consider the rate equation for the density of charge defect, which is given by
\begin{eqnarray}
	\label{eq:rate-eq}
    	\frac{dn_{3}}{dt} =  A_1 \left(e^{-8\beta J} n_1^{2} - 9 n_{3}^2\right)   +  A_2 \left(e^{-4\beta J} n_1^{2} - 3 n_1 n_{3} \right), \qquad
\end{eqnarray}
The two coefficients $A_1,~A_2$ are temperature-independent parameters to be determined by fitting with the MC simulation results. Starting from an initially random configuration where $n_3(0) = 1/4$, the above rate equation is numerically integrated using a variable step and variable order ordinary differential equation solver.  By fitting the integration results with data points from one particular cooling time $\tau_Q = 160$ of the linear schedule, we obtained $A_1=1.566$ and $A_2=0.600$. Remarkably, as shown in FIG.~\ref{fig:nd-cooling},  the rate equation description based on these two fitted parameters gives consistent results with the Glauber dynamics simulations for all three algebraic cooling schedules and a wide range of cooling time $\tau_Q$. Moreover, excellent agreements were obtained not only for the residual defect density at the end of cooling, but also for the entire relaxation process.

\section{Asymptotic solution of the rate equation}
\label{sec:sol}

In this section, we solve the rate equation~(\ref{eq:rate-eq}) in the slow cooling limit $\tau_Q \gg \tau$, where $\tau$ is the nearly constant relaxation time in Eq.~(\ref{eq:tau}). 
At low temperatures, the defect density is small $n_3 \ll 1$, and we can approximate $n_1 = 1 - n_3 \approx 1$. This also means that the pair annihilation term in the rate equation can be neglected. Moreover, spontaneous pair creation of charge defects with an exponential factor $e^{-8\beta}$ is also negligible compared with the generation of a single defect. This means that the $A_1$ term, corresponding to pair creation/annihilation processes shown in FIG.~\ref{fig:reactions}(d), can be neglected. The rate equation then becomes
\begin{eqnarray}
	\label{eq:dn3-lowT}
	\frac{dn_3}{dt} = -\frac{n_3}{\tau} + \frac{1}{3\tau} e^{-4 \beta(t) J},
\end{eqnarray}
where the relaxation time $\tau = 1/(3 A_2)$. Using numerical value $A_2 = 0.600$ fitted from the cooling simulations, we obtain $\tau \approx 0.555$, which is consistent with the estimate from the instant quench simulations shown in FIG.~\ref{fig:relaxation}. The above equation can be readily integrated to give
\begin{eqnarray}
	& & n_3(t) = n_3(t_0) e^{-(t-t_0)/\tau} \\ \nonumber 
	& & \qquad + \frac{1}{3\tau} \int_{t_0}^t e^{-4\beta(s) J} e^{-(t-s)/\tau} ds.
\end{eqnarray}
Here $t_0$ is a characteristic instant after which the low-$T$ approximation Eq.~(\ref{eq:dn3-lowT}) of the rate equation is valid.
The residual charge defect is given by $n_3(t = \tau_Q)$. In the slow cooling limit $\tau_Q \gg \tau$, the first initial value term with $n_3(t_0)$ tends to zero. We have
\begin{eqnarray}
	n_3(\tau_Q) = \frac{1}{3\tau} \int_{t_0}^{\tau_Q} e^{-4\beta(s) J} e^{-(\tau_Q-s)/\tau} ds
\end{eqnarray}
This integral in the slow cooling limit is dominated by the time region $s \lesssim \tau_Q$ because of the exponentially decaying memory function $\exp[-(\tau_Q - s)/\tau]$. For convenience, we introduce a change of variable $\eta = (\tau_Q - s)/\tau$. By expressing the Arrhenius factor $e^{-4\beta J}$ in terms of the dimensionless $\gamma$, we have
\begin{eqnarray}
	n_3(\tau_Q) = \frac{1}{3} \int_0^{(\tau_Q - t_0)/\tau} \frac{1 - \gamma(\eta)}{1 + \gamma(\eta)} e^{-\eta} \, d\eta,
\end{eqnarray}
The integral is now dominated by $\eta \gtrsim 0$, where $\gamma \approx 1$. Approximating the denominator $1+\gamma \approx 2$, and substituting Eq.~(\ref{eq:gamma}) for $\gamma(\eta)$ in the numerator, we obtain
\begin{eqnarray}
	n_3(\tau_Q) = \frac{1}{6}\left(\frac{\tau}{\tau_Q}\right)^\alpha \int_0^{(\tau_Q - t_0)/\tau} \eta^\alpha e^{-\eta} \, d\eta.
\end{eqnarray}
The upper limit of the integral can be replaced by $\infty$ in the $\tau_Q \gg \tau$ limit, the defect density at the end of cooling is then given by
\begin{eqnarray}
	\label{eq:n3-analytical}
	n_3(\tau_Q) = \frac{\Gamma(1+\alpha)}{6}\left(\frac{\tau}{\tau_Q}\right)^\alpha,
\end{eqnarray}
where $\Gamma(x)$ is the Gamma function. Importantly, here we show that the residual charge defects exhibit a scaling relation with the cooling rate with an exponent controlled by the exponent $\alpha$ of the cooling schedule. This result is also completely consistent with the power-law behavior obtained in our Glauber dynamics simulations, as shown by the dashed lines in FIG.~\ref{fig:nd-cooling}. 

In fact, the above argument can be applied to cooling schedules where the exponential factor admits a power law expansion: $e^{-4\beta(t) J} = a (\tau_Q - t)^\alpha + \cdots$ at $t \lesssim \tau_Q$. It is worth noting that, since either the Glauber or Metropolis dynamics for Ising spins is controlled by this Arrhenius factor, it is more natural to define cooling schedules in terms of this factor or equivalently the dimensionless parameter Eq.~(\ref{eq:gm-factor}).  The above series expansion of $e^{-4\beta J}$ corresponds to a physical temperature which vanishes in such a way that its inverse diverges logarithmically near~$t = \tau_Q$:
\begin{eqnarray}
	T(t) \approx \frac{4 J}{\alpha \bigl| \log(\tau_Q - t) \bigr|}.
\end{eqnarray}
Finally, we note that that non-universal behaviors, in particular non-power-law dependences, are expected for general cooling schedules that do not belong to this class. For example, for the linear in $T$ cooling schedule: $T(t) \sim (1-t/\tau_Q)$, which belongs to a special case of the so-called exponential cooling protocol, our Glauber dynamics simulation find a residual charge defect density which decays exponentially with cooling time: $n_3(\tau_Q) \sim \exp(-{\rm const.} \times \tau_Q)$. This result is consistent with the non-power-law behavior observed in the quench simulations of artificial colloidal kagome ice~\cite{libal20}.

\section{summary and outlook}
\label{sec:summary}

To summarize, we have presented a comprehensive study on the quench dynamics of nearest-neighbor kagome spin ice. This highly frustrated spin system is not only an important model in statistical physics and a representative system of geometrical frustration. The kagome spin ice exhibits an unusual non-topological defects, corresponding to triangle-simplexes which violate the ice-rule constrants, in the low temperature ice phase. The relaxation time results from the decay of such non-topological solitons remains finite and nearly temperature independent. As a result, the standard KZ mechanism cannot be applied to the describe the quench dynamics of kagome ice. Yet, our extensive simulations and rate-equation analysis demonstrate that, for a special case of algebraic cooling protocols, the residual defect density in kagome spin ice exhibits a power-law relation with the cooling rate. We further show that the power-law dependences result from accumulation of excess defects during the non-adiabatic evolution at the late stage of cooling. Our analysis can also be extended to the quench dynamics of other systems with non-topological solitons.

The kagome spin ice considered in this work is an idealized frustrated Ising system with nearest-neighbor interactions. In realistic versions of kagome spin ice, either in magnetic compounds such as Dy$_3$Mg$_2$Sb$_3$O$_{14}$~\cite{paddison16,dun16} or in artificial nanomagnetic realizations~\cite{tanaka06,qi08,yue22,zhang13,farhan16}, the magnetic Coulomb interaction between the background $Q=\pm 1$ magnetic charges leads to an intriguing intermediate phase where the charge degrees of freedom develop a long-range order, yet spins remain disordered~\cite{moller09,chern11}. The Coulomb interaction is minimized by a staggered charge distribution with all up-triangles assume, e.g. $Q = +1$, while all down-triangles are in the $Q = -1$ state, or vice versa. This charge ordered state can thus be described by an Ising order parameter $m = \langle Q_{\triangle} - Q_{\bigtriangledown} \rangle$. Importantly, phase transition into this charge ordered state thus belongs to the 2D Ising universality class~\cite{moller09,chern11}. 
As temperature is further lowered, a long-range magnetic order with a tripled unit cell is eventually induced by the long-range dipole-dipole interaction. The corresponding magnetic transition is of the 2D three-state Potts universality class~\cite{chern11}. Nonequilibrium critical dynamics due to thermal quenches across these two phase transitions is expected to be well described by the KZ mechanism.  A detailed investigation of kagome spin ice with dipolar interactions will be left for future study. 

Finally, we note that a rather different scenario occurs in the 3D pyrochlore spin ice, and its 2D counterpart the checkerboard spin ice. In both cases, the lattices are composed of simplexes (e.g. tetrahedra in the pyrochlore lattice) with four Ising spins. As a result, the ground-state manifold comprises 2-in-2out simplexes with zero magnetic charge. Importantly, elementary excitations of such ice phases, i.e. simplexes that violates the 2-in-2-out ice rules, are topological in nature and carry a net magnetic charge, effectively behaving as emergent magnetic monopoles. Also contrary to the ice phase of kagome, both pyrochlore and checkerboard spin ices approaches a critical state as $T \to 0$, similar to the 1D Ising chain (however, unlike 1D Ising chain, spins remain disordered at the $T=0$ critical state in spin ice). This critical point in spin ices is unconventional as the correlation length diverges exponentially, instead of algebraically, when approaching zero temperature. Nonetheless, the KZ mechanism can be generalized to understand the nonequilibrium generation of topological defects when such spin ices are quenched to the critical point. A detailed account of the quench dynamics of pyrochlore spin ice will be presented elsewhere~\cite{fan23}.

\begin{acknowledgments}
The authors thank Adolfo del Campo for useful discussions and collaboration on related projects. The work was supported by the US Department of Energy Basic Energy Sciences under Contract No. DE-SC0020330. The authors also acknowledge the support of Research Computing at the University of Virginia. 
\end{acknowledgments}

\end{document}